\documentstyle[aps,amssymb,amsfonts,times,mathptm,12pt]{revtex}
\begin{document}
\small
\normalsize
\protect\newtheorem{principle}{Principle} 
\protect\newtheorem{theo}[principle]{Theorem}
\protect\newtheorem{prop}[principle]{Proposition}
\protect\newtheorem{lem}[principle]{Lemma}
\protect\newtheorem{co}[principle]{Corollary}
\protect\newtheorem{de}[principle]{Definition}
\newtheorem{ex}[principle]{Example}
\newtheorem{rema}[principle]{Remark}
\renewcommand{\baselinestretch}{1}
\small
\normalsize
\title{A separability criterion for density operators}
\author{Oliver Rudolph \thanks{email: rudolph@starlab.net}}
\address{Starlab nv/sa, Physics Division, Boulevard Saint Michel
47, B-1040 Brussels, Belgium}
\maketitle
\begin{abstract}
\noindent We give a necessary and sufficient condition for a mixed
quantum mechanical state to be separable. The criterion is formulated
as a boundedness condition in terms of the greatest cross
norm on the tensor product of trace class operators.
\end{abstract}
\section{Introduction}
The question of separability of density operators on finite dimensional
Hilbert
spaces has recently been studied extensively
in the context of quantum information
theory, see, e.g.,
\cite{Horodecki97,Kraus99,Peres96,PittengerR99,PittengerR00,Rungta00}
and references therein.
In the present work we provide a simple mathematical characterisation
of separable density operators.

Throughout this paper the set of trace class operators on some Hilbert
space ${\cal H}$ is denoted by ${\cal T}({\cal H})$ and
the set of bounded operators on ${\cal H}$ is denoted by
${\cal B}({\cal H})$. A density operator is a positive trace
class operator with trace one.

\begin{de}
Let ${\cal H}_1$ and ${\cal H}_2$ be two Hilbert spaces of
arbitrary dimension. A density operator $\varrho$ on the tensor product
${\cal H}_1
\otimes {\cal H}_2$ is called \emph{separable} if there exist a
family $\left\{ \omega_{i} \right\}$ of positive real numbers, a family
$\left\{ \rho^{(1)}_i \right\}$ of density operators on
${\cal H}_1$ and a family $\left\{ \rho^{(2)}_i \right\}$ of
density operators
on ${\cal H}_2$ such that
\begin{equation} \label{e1}
\varrho = \sum_{i} \omega_{i} \rho^{(1)}_i \otimes \rho^{(2)}_i,
\end{equation}
where the sum converges in trace class norm.
\end{de}
Consider the spaces ${\cal T}({\cal H}_1)$ and ${\cal T}({\cal H}_2)$
of trace class operators on ${\cal H}_1$ and ${\cal H}_2$
respectively. Both spaces are Banach spaces when equipped with the trace
class norm $\Vert \cdot \Vert_1^{(1)}$ or $\Vert \cdot
\Vert_1^{(2)}$ respectively, see, e.g., Schatten \cite{Schatten70}. In
the sequel we shall drop the superscript and write $\Vert \cdot
\Vert_1$ for both norms, slightly abusing the notation; it will be
always clear from the context which norm is meant.
The algebraic tensor product ${\cal T}({\cal H}_1)
\otimes_{\rm alg} {\cal T}({\cal H}_2)$
of ${\cal T}({\cal H}_1)$
and ${\cal T}({\cal H}_2)$ is defined as the set of all finite
linear combinations of elementary tensors $u \otimes
{v}$, i.e., the set of all finite sums $\sum_{i=1}^n u_i \otimes
{v}_i$ where $u_i \in {\cal T}({\cal H}_1)$ and
${v}_i \in {\cal T}({\cal H}_2)$ for all $i$.

It is well known that we can define a cross norm on ${\cal T}({\cal H}_1)
\otimes_{\rm alg} {\cal T}({\cal H}_2)$ by \cite{WeggeOlsen93}
\begin{equation}
\Vert t \Vert_\gamma := \inf \left\{ \sum_{i=1}^n
\left\Vert u_i \right\Vert_1 \, \left\Vert
{v}_i \right\Vert_1 \, \left\vert \, t = \sum_{i=1}^n u_i
\otimes {v}_i \right. \right\}, \end{equation} where $t \in
{\cal T}({\cal H}_1)
\otimes_{\rm alg} {\cal T}({\cal H}_2)$ and where the infimum runs over
all finite decompositions of $t$ into elementary tensors.
It is well known that $\Vert \cdot \Vert_\gamma$ majorizes any
subcross seminorm on ${\cal T}({\cal H}_1)
\otimes_{\rm alg} {\cal T}({\cal H}_2)$ and that the completion
${\cal T}({\cal H}_1)
\otimes_{\gamma} {\cal T}({\cal H}_2)$ of
${\cal T}({\cal H}_1)
\otimes_{\rm alg} {\cal T}({\cal H}_2)$ with respect to
$\Vert \cdot \Vert_\gamma$ is a Banach algebra \cite{WeggeOlsen93}.

In the following we
specialize to the situation where both ${\cal H}_1$ and ${\cal
H}_2$ are finite dimensional, hence ${\cal T}({\cal H}_1) =
{\cal B}({\cal H}_1)$ and ${\cal T}({\cal H}_2) =
{\cal B}({\cal H}_2)$. It is well known that the completion of
${\cal B}({\cal H}_1)
\otimes_{\rm alg} {\cal B}({\cal H}_2)$ with respect to the spatial norm
on ${\cal B}({\cal H}_1)
\otimes_{\rm alg} {\cal B}({\cal H}_2)$
is equal to ${\cal B}({\cal H}_1 \otimes {\cal H}_2)$, see,
e.g., \cite{KadisonR86}, Example 11.1.6. (The spatial norm
on ${\cal B}({\cal H}_1)
\otimes_{\rm alg} {\cal B}({\cal H}_2)$ is the
norm induced by the operator
norm on ${\cal B}({\cal H}_1 \otimes {\cal H}_2)$.)
As ${\cal H}_1$ and ${\cal
H}_2$ are finite dimensional, both ${\cal T}({\cal H}_1) =
{\cal B}({\cal H}_1)$ and ${\cal T}({\cal H}_2) =
{\cal B}({\cal H}_2)$ are nuclear. By nuclearity it follows that
the completion of ${\cal B}({\cal H}_1)
\otimes_{\rm alg} {\cal B}({\cal H}_2)$ with respect to $\Vert \cdot
\Vert_\gamma$, denoted by
${\cal B}({\cal H}_1)
\otimes_{\gamma} {\cal B}({\cal H}_2)$,
equals ${\cal B}({\cal H}_1 \otimes {\cal H}_2)$.
Moreover in finite
dimensions all Banach space norms on ${\cal B}({\cal H}_1
\otimes {\cal H}_2)$, in particular
the operator norm $\Vert \cdot \Vert$, the trace class norm
$\Vert \cdot \Vert_1$, and the norm $\Vert \cdot
\Vert_\gamma$, are
equivalent, i.e., generate the same topology on ${\cal
B}({\cal H}_1 \otimes {\cal H}_2)$.
\section{The separability criterion}
\begin{lem}
Let ${\cal H}_1$ and ${\cal H}_2$ be Hilbert
spaces, and $\varrho$ be a density operator on ${\cal H}_1 \otimes
{\cal H}_2$. Then $\Vert \varrho \Vert_\gamma \leq 1$ if and only
if $\Vert \varrho \Vert_\gamma =1$.
\end{lem}
\emph{Proof}: This follows from $1 = \Vert \varrho \Vert_1 \leq
\Vert \varrho \Vert_\gamma.$ $\Box$ \\
\begin{prop} \label{p4}
Let ${\cal H}_1$ and ${\cal H}_2$ be finite dimensional Hilbert
spaces and let $\varrho$ be a separable density operator
on ${\cal H}_1 \otimes {\cal H}_2$, then $\Vert \varrho \Vert_\gamma
\leq 1$. \end{prop}
\emph{Proof}: Let $\varrho$ be a separable density operator on
${\cal H}_1 \otimes {\cal H}_2$, then there exist a family
$\{ \omega_{i} \}$ of positive real  numbers, a family
$\left\{ \rho^{(1)}_i \right\}$ of density operators on
${\cal H}_1$ and a family $\left\{ \rho^{(2)}_i \right\}$
of density operators
on ${\cal H}_2$ such that
\[ \varrho = \sum_{i} \omega_{i} \rho^{(1)}_i \otimes \rho^{(2)}_i, \]
where the sum converges in trace class norm. If this sum is finite,
then obviously $\Vert \varrho \Vert_\gamma \leq 1$. If the sum is infinite,
consider the sequence
$\{ \varrho_n \}$ of trace class operators where $\varrho_n \equiv
\sum_{i=1}^n \omega_{i} \rho^{(1)}_i \otimes \rho^{(2)}_i$. The
sequence $\{ \varrho_n \}$ converges to $\varrho$ in trace class
norm and is a Cauchy sequence with respect to $\Vert \cdot \Vert_\gamma$.
Thus $\{ \varrho_n \}$ converges to $\varrho$ with respect to the
norm $\Vert \cdot \Vert_\gamma$
and we have $\Vert \varrho_n \Vert_\gamma \leq 1$ for all $n$.
As $\Vert \varrho \Vert_\gamma \leq \Vert \varrho - \varrho_n \Vert_\gamma
+ \Vert \varrho_n \Vert_\gamma$ for all $n$,
also $\Vert \varrho \Vert_\gamma \leq 1$. $\Box$ \\ \\
All density operators $\varrho$ satisfy \[ 1 = {\rm tr}(\varrho) =
\Vert \varrho \Vert_1 \leq \Vert \varrho \Vert_\gamma \] with
equality if $\varrho$ is separable. Thus one might tentatively consider
the difference $\Vert \varrho \Vert_\gamma -
\Vert \varrho \Vert_1$ as a measure of nonseparability.
\begin{prop} \label{p5}
Let ${\cal H}_1$ and ${\cal H}_2$ be finite dimensional Hilbert
spaces and let $\varrho$ be a density operator on ${\cal H}_1
\otimes {\cal H}_2$ with
$\Vert \varrho \Vert_\gamma \leq 1$, then $\varrho$ is
separable. \end{prop}
\emph{Proof}: Let $\varrho$ be a density operator on ${\cal H}_1 \otimes
{\cal H}_2$ with $\Vert \varrho \Vert_\gamma \leq 1$.
We divide the proof of separability into two steps. Firstly we show that
for every $\delta > 0$
there exist families $\{ x_i(\delta) \}$ and $\{ y_i(\delta) \}$
of trace class
operators on ${\cal H}_1$ and ${\cal H}_2$ respectively such that $\varrho
= \sum_i x_i(\delta) \otimes y_i(\delta)$,
where the sum converges with respect to
the trace class norm, and such that \[ \sum_i \Vert x_i(\delta)
\Vert_1 \Vert
y_i(\delta) \Vert_1 \leq \Vert \varrho \Vert_1 + \delta = 1 + \delta. \]
As
$\varrho \in {\cal B}({\cal H}_1) \otimes_\gamma {\cal B}({\cal
H}_2)$, there exist elements $\varrho_n(\delta) \in {\cal B}({\cal H}_1)
\otimes_{\rm alg} {\cal B}({\cal H}_2)$, where $n \in {\Bbb N}$,
such that \[
\Vert \varrho - \varrho_n(\delta) \Vert_\gamma < \frac{1}{2^{n+3}}
\delta. \]
Consequently $\Vert \varrho_{n+1}(\delta) - \varrho_n(\delta)
\Vert_\gamma <
\frac{1}{2^{n+2}} \delta$ for all $n$. Therefore $\varrho_{n+1}(\delta) -
\varrho_n(\delta)$ can be written in the form \[
\varrho_{n+1}(\delta)  - \varrho_n(\delta) =
\sum_{k_{n+1}=1}^{m_{n+1}} x_{k_{n+1}}^{(n+1)}(\delta) \otimes
y_{k_{n+1}}^{(n+1)}(\delta) \]
with $x_{k_{n+1}}^{(n+1)}(\delta) \in {\cal B}({\cal H}_1)$,
$y_{k_{n+1}}^{(n+1)}(\delta) \in
{\cal B}({\cal H}_2)$ and \[ \sum_{k_{n+1}=1}^{m_{n+1}}
\left\Vert x_{k_{n+1}}^{(n+1)}(\delta)
\right\Vert_1
\left\Vert y_{k_{n+1}}^{(n+1)}(\delta) \right\Vert_1
\leq \frac{1}{2^{n+2}} \delta. \]
Since \[ \Vert \varrho_0(\delta) \Vert_\gamma
\leq \Vert \varrho \Vert_\gamma +
\Vert \varrho_0(\delta) - \varrho \Vert_\gamma <
\Vert \varrho \Vert_\gamma +
\frac{1}{2} \delta, \] $\varrho_0(\delta)$ can be represented as
\[ \varrho_0(\delta) =
\sum_{k_0=1}^{m_0} x_{k_0}^{(0)}(\delta) \otimes y_{k_0}^{(0)}(\delta) \]
with $x_{k_0}^{(0)}(\delta) \in {\cal B}({\cal H}_1)$,
$y_{k_0}^{(0)}(\delta) \in {\cal
B}({\cal H}_2)$ and \[ \sum_{k_0=1}^{m_0} \left\Vert x_{k_0}^{(0)}(\delta)
\right\Vert_1
\left\Vert y_{k_0}^{(0)}(\delta) \right\Vert_1
\leq \Vert \varrho \Vert_\gamma + \frac{1}{2}
\delta. \] Consequently,
\begin{eqnarray} \varrho & = & \varrho_0(\delta) + \sum_{n \in {\Bbb N}}
\left( \varrho_{n+1}(\delta) - \varrho_n(\delta) \right)
\label{eq8} \\ & = & \sum_{n \in {\Bbb N}}
\sum_{k_n=1}^{m_n} x_{k_n}^{(n)}(\delta) \otimes y_{k_n}^{(n)}(\delta).
\label{eq9} \end{eqnarray} Thus we arrive at
\begin{equation} \label{eq0} 1 = \Vert \varrho \Vert_1 \leq
\sum_{n \in {\Bbb N}}
\sum_{k_n=1}^{m_n} \left\Vert x_{k_n}^{(n)}(\delta)
\right\Vert_1 \left\Vert
y_{k_n}^{(n)}(\delta) \right\Vert_1 \leq
\Vert \varrho \Vert_\gamma + \delta \leq 1 + \delta \end{equation}
which concludes the first part of
our proof. By virtue of (\ref{eq0}) the
sequence $\left\{x_{k_n}^{(n)}(\frac{1}{N})
\otimes y_{k_n}^{(n)}(\frac{1}{N}) \right\}_{N \in {\Bbb N} \backslash 0}$
is bounded with respect to
the trace class norm for every $n,{k_n}$. Therefore, by
possibly passing to a subsequence, we can assume that
$\left\{x_{k_n}^{(n)}(\frac{1}{N})
\otimes y_{k_n}^{(n)}(\frac{1}{N}) \right\}_N$ converges in trace class norm
to a trace class operator $x_{k_n}^{(n)} \otimes y_{k_n}^{(n)}$
for $N \to \infty$.
From (\ref{eq0}) we infer that \[ \left\Vert \sum_{n} \sum_{k_n}
x_{k_n}^{(n)}
\otimes y_{k_n}^{(n)} \right\Vert_1 \leq \sum_{n} \sum_{k_n} \left\Vert
x_{k_n}^{(n)}
\right\Vert_1 \left\Vert y_{k_n}^{(n)} \right\Vert_1 =1, \] and thus
$\sum_n \sum_{k_n} x_{k_n}^{(n)} \otimes y_{k_n}^{(n)}$ is convergent.

If we let $\delta \in ]0,1]$, then
$\Vert \varrho_{n+1}(\delta) - \varrho_n(\delta) \Vert_\gamma <
\frac{1}{2^{n+2}} \delta \leq \frac{1}{2^{n+2}}$ and $\Vert
\varrho_0(\delta) \Vert_\gamma < \Vert \varrho \Vert_\gamma + \frac{1}{2}
\delta \leq \Vert \varrho \Vert_\gamma + \frac{1}{2}$.
Thus we find that
\[ \sup_\delta \Vert \varrho_0(\delta) \Vert_\gamma + \sum_n
\sup_\delta \Vert \varrho_{n+1}(\delta) - \varrho_n(\delta) \Vert_\gamma
< \infty. \]
Thus we conclude (Weierstra\ss{} convergence criterion) that the
series (\ref{eq8}) converges uniformly on $]0,1]$ and therefore we can
interchange the infinite sums in (\ref{eq8}) and (\ref{eq9}) with
the limit $N \to \infty$, arriving at
\[ \varrho = \lim_{N \to \infty} \sum_n \sum_{k_n} x_{k_n}^{(n)}
\left( 1/N \right) \otimes
y_{k_n}^{(n)} \left( 1/N \right) = \sum_n \sum_{k_n} \lim_{N \to \infty}
\left(x_{k_n}^{(n)} \left( 1/N \right) \otimes y_{k_n}^{(n)} \left(
1/N \right) \right)
= \sum_n \sum_{k_n} x_{k_n}^{(n)} \otimes y_{k_n}^{(n)}. \]
As moreover, by (\ref{eq0}),
\begin{eqnarray*}
1  = \left\vert {\rm tr}(\varrho) \right\vert
& = & \left\vert \sum_n \sum_{k_n} {\rm tr}
\left(x_{k_n}^{(n)}(\delta) \right) \otimes
{\rm tr} \left(y_{k_n}^{(n)}(\delta) \right) \right\vert \\
& \leq & \sum_n \sum_{k_n}
\left\vert {\rm tr} \left( x_{k_n}^{(n)}(\delta) \right)
{\rm tr} \left( y_{k_n}^{(n)}(\delta) \right) \right\vert \\ & \leq &
\sum_n \sum_{k_n} \left\Vert x_{k_n}^{(n)}(\delta) \right\Vert_1
\left\Vert y_{k_n}^{(n)}(\delta) \right\Vert_1 \\ & \leq &
\sum_n \sum_{k_n}
\left\Vert x_{k_n}^{(n)} \right\Vert_1
\left\Vert y_{k_n}^{(n)} \right\Vert_1 +\delta \\ & = & 1 + \delta,
\end{eqnarray*}
we see that $\left\vert {\rm tr} \left(x_{k_n}^{(n)}(\delta) \right) \otimes
{\rm tr} \left(y_{k_n}^{(n)}(\delta) \right) \right\vert$ converges to
$\left\Vert x_{k_n}^{(n)} \right\Vert_1
\left\Vert y_{k_n}^{(n)} \right\Vert_1$ for all ${k_n},n$.
Thus $\left\Vert x_{k_n}^{(n)} \right\Vert_1
\left\Vert y_{k_n}^{(n)} \right\Vert_1 =
\left\vert {\rm tr} \left(x_{k_n}^{(n)} \right) \right\vert
\left\vert {\rm tr} \left(y_{k_n}^{(n)} \right) \right\vert$
and therefore $\left\Vert x_{k_n}^{(n)} \right\Vert_1
= \left\vert {\rm tr} \left(x_{k_n}^{(n)} \right) \right\vert$
and $\left\Vert y_{k_n}^{(n)} \right\Vert_1 =
\left\vert {\rm tr} \left(y_{k_n}^{(n)} \right) \right\vert$.
This implies that we can choose all $x_{k_n}^{(n)}$ and
$y_{k_n}^{(n)}$ as positive
trace class operators. This proves that $\varrho$ is separable.
$\Box$ \\ \\
Putting all our results together we arrive at the main theorem of
this paper
\begin{theo} \label{t1}
Let ${\cal H}_1$ and ${\cal H}_2$ be finite dimensional Hilbert
spaces and $\varrho$ be a density operator on ${\cal H}_1
\otimes {\cal H}_2$. Then $\varrho$ is separable if and only if
$\Vert \varrho \Vert_\gamma = 1$. \end{theo}

\section{Conclusion}
To conclude we have been able to prove a new mathematical
separability criterion for density operators: a density operator $\varrho$
on a finite dimensional tensor product Hilbert space is separable
if and only if $\Vert \varrho \Vert_\gamma = 1$.
Our results also imply that the difference $\Vert \varrho \Vert_\gamma
- \Vert \varrho \Vert_1 = \Vert \varrho \Vert_\gamma -1$ may be
considered as a quantitative measure of entanglement.
In general it will be difficult to compute $\Vert \varrho
\Vert_\gamma$ exactly, and accordingly Theorem \ref{t1} is
unlikely to provide a practical tool to decide whether or not a given
density operator is separable without explicitly constructing a
representation of the form (\ref{e1}).
However, Theorem \ref{t1} provides some principal
insight into the structure of
the space of density operators and therefore is of some interest
in it's own right.
We have restricted ourselves to density operators on a tensor
product Hilbert space of two finite dimensional Hilbert spaces. It
is straightforward, however, to generalize our results to the
situation of density operators defined on a tensor product of more
than two, but at most finitely many, finite dimensional Hilbert
spaces.


\begin{thebibliography}{99}
\bibitem{Horodecki97} {Horodecki M, Horodecki P and Horodecki R 1996}
{Separability of mixed states: necessary and sufficient
conditions} \emph{Phys.~Lett.~A} \textbf{78} 1-8.
\bibitem{Kraus99} Kraus B, Cirac J I, Karnas S and Lewenstein
M 1999 {Separability in $2 \times N$ composite quantum systems}
\emph{Preprint, quant-ph/9912010}.
\bibitem{Peres96} Peres A 1996 Separability criterion for
density matrices \emph{Phys.~Rev.~Lett}.~\textbf{77} 1413-1415.
\bibitem{PittengerR99} Pittenger A O and Rubin M H 1999 {Complete
separability and Fourier representations of n-qubit states} \emph{Preprint
quant-ph/9912116.}
\bibitem{PittengerR00} Pittenger A O and Rubin M H 2000 {Separability
and Fourier representations of density matrices} \emph{Preprint
quant-ph/0001014.}
\bibitem{Rungta00} Rungta P, Munro W J, Nemoto K, Deuar P,
Milburn G J and Caves C M 2000 {Qudit entanglement} \emph{Preprint
quant-ph/0001075.}
\bibitem{Schatten70} Schatten R 1970 \emph{Norm Ideals of
Completely Continuous
Operators} 2nd edn.~(Berlin: Springer).
\bibitem{WeggeOlsen93} {Wegge-Olsen N E 1993} \emph{K-Theory
and $C^*$-algebras} (Oxford: Oxford University Press).
\bibitem{KadisonR86} {Kadison R V and Ringrose J R 1983 \& 1986}
\emph{Fundamentals of the Theory of Operator Algebras} vol I \&
II (Orlando: Academic).
\end{thebibliography}
\end{document}